# MODELING AND ANALYSIS OF MULTIPLE ELECTROSTATIC ACTUATORS ON THE RESPONSE OF VIBROTACTILE HAPTIC DEVICE

**Santosh Mohan Rajkumar[1], Kumar Vikram Singh[1], Jeong-Hoi Koo[1]**

[1]Department of Mechanical & Manufacturing Engineering, Miami University, Oxford, Ohio, USA

**ABSTRACT**

*In this research, modeling, and analysis of a beam-type touchscreen interface with multiple actuators is considered. A mechanical model of a touch screen system, as thin beams, is developed with embedded electrostatic actuators at different spatial locations. This discrete finite element-based model is developed to compute the analytical and numerical vibrotactile response due to multiple actuators excited with varying frequency and amplitude. The model is tested with spring-damper boundary conditions incorporating sinusoidal excitations in the human haptic range. An analytical solution is proposed to obtain the vibrotactile response of the touch surface for different frequencies of excitations, the number of actuators, actuator stiffness, and actuator positions. The effect of the mechanical properties of the touch surface on vibrotactile feedback provided to the user feedback is explored. Investigation of optimal location and number of actuators for a desired localized response, such as the magnitude of acceleration and variation in acceleration response for a desired zone on the interface, is carried out. It has been shown that a wide variety of localizable vibrotactile feedback can be generated on the touch surface using different frequencies of excitations, different actuator stiffness, number of actuators, and actuator positions. Having a mechanical model will facilitate simulation studies capable of incorporating more testing scenarios that may not be feasible to physically test.*

## NOMENCLATURE

| | |
|---|---|
| ERA | Electrostatic Resonating Actuator |
| $k_b$ | stiffness of an ERA |
| $m_b$ | mass of bolt connecting touch surface to ERA |
| $c$ | damping co-efficient of an ERA |
| $a$ | maximum displacement of an ERA |
| $L$ | length of the touch surface |
| $b$ | width of the touch surface |
| $h$ | thickness of the touch surface |
| $E$ | elastic modulus |
| $I$ | moment of inertia |
| $A$ | cross-sectional area |
| $\rho$ | density |
| $u$ | transverse displacement |
| $\mathbf{d}$ | vector of nodal displacements and slopes of the touch bar display |
| $x$ | spatial location |
| $t$ | time instant |

## INTRODUCTION

Humans leverage haptics or the sense of touch to perform a wide variety of exploratory and manipulatory tasks. Haptic technologies find application in robotics, telesurgery, biomedical engineering, interactive computing, consumer electronic devices, digital musical instruments, etc. [1, 2]. Tactile sensation is a type of haptic sensation sensed by the mechanoreceptors of the human skin. The tactile sensation can be based on pressure, shear, vibration, etc. [2]. Vibrotactile feedback is a coveted feature for digital touch screen devices for an effective, precise, and captivating user experience. Humans are most sensitive to vibrotactile sensations in the range 150-250 Hz [3]. Haptic modules currently in place can generate effective vibrotactile feedback in devices with smaller touch screens [3]. The tactile sensation diminishes in strength for larger touchscreen displays as surface area increases [4]. Large touch screen displays find applications in a variety of areas such as digital musical instruments (DMIs), information kiosks, airports, restaurant chains, tablet computers, touch screen laptops / personal computers, vehicle infotainment systems, etc. [1, 5]. However, vibrotactile feedback in large touchscreen devices is mostly absent owing to the lack of availability of actuators capable of generating vibration magnitudes for larger screens [3,4]. Moreover, with a given actuator capable of creating vibrotactile feedback on large touch screens, there is still a challenge to generate vibrations at specific places (localized vibrotactile feedback) on the screen. Also, vibrotactile feedback localization for large multi-touch displays is challenging as it is difficult to control the wave of vibration propagating through the surface [6]. The localized vibrotactile sensation in large touch screen systems depends on the type and number of actuators used and their position on the touch bar or plate interface. More than one actuator capable of exciting the interface with varying amplitude and frequency may be needed for targeted vibrotactile feedback at a desired location on the interface.

Various types of actuators are available for vibrotactile feedback generation on haptic surfaces. However, haptic actuators capable of producing adequate vibrotactile feedback are not abundant [7]. Haptic actuators can be broadly classified into five types [6, 8]: 1) eccentric rotary mass (ERM) actuators, 2) linear resonant actuators (LRA), 3) piezoelectric actuators, 4) controllable fluid (such as magnetorheological or MR fluid and electrorheological or ER fluid) actuators, and 5) electrostatic actuators. ERM actuators can create large relative vibrations but with constant displacement. LRA actuators are compact with a very narrow frequency bandwidth of vibrations. Piezoelectric actuators are limited by very high input voltage requirements and cost. MR fluid actuators are limited by the size of their magnets, and they can't be used in capacitive touch screens. Electrostatic actuators can be made in small and thin modules [7], making them desirable candidates for touchscreen device applications.

Several attempts have been made in the literature to generate vibrotactile sensations in large-screen devices. Mason et al. [3] experimentally demonstrated the feasibility of localized vibrotactile feedback generation with a limited number of electrostatic resonating actuators (ERA) for bar/ beam type display interfaces. With a new touch bar design, it has been shown experimentally that localizable vibrotactile feedback can be generated by varying excitation frequencies, amplitude, and duration. However, a comprehensive modeling and analysis paradigm is required to understand the effect of actuator parameters, touch surface properties, number of excitations needed, positions of excitation, and excitation parameters. Also, in obtaining a desired vibrotactile pattern on the touch surface, modeling can help determine required excitations, their positions, and actuator properties. Jansen et al. [8] proposed a resistive touch surface capable of multi-touch localizable haptic feedback. The proposed system can't be applied to capacitive touch screens due to the presence of metallic particles in MR fluid and is constrained by the size of the magnets used. Bai et al. [9] use time reversal wave focusing method with electromagnetic actuators to create a vibration pulse on a size of 2 cm diameter on a large surface. Hudin et al. [10] also used time-reversal wave focusing with piezoelectric actuators to generate a pulse vibration on a size of 0.54 cm on a large plate. The time reversal wave focusing technique produces vibrotactile sensation in the supersonic frequency range, and the resolution of haptic sensation is on the higher side. Woo et al. [11] implemented eigenfunction superposition and traveling waves to create localized vibrotactile feedback on a large touchscreen device. The resolution of vibration localization in the eigenfunction superposition method is compromised. The generation of traveling waves to replicate a target vibration pattern on a surface is not effective for long-duration tactile feedback generation. Both the eigenfunction and traveling wave methods use many actuators, making them less germane to compact, size-constrained applications. In Emgin et al. [12], vibration localization at 84 grid points of a large touch surface was achieved using a predetermined vibration lookup table. The resolution of vibrotactile sensation is high in the study. Van Duong et al. [13] delineated localized vibrations using a relaxor ferromagnetic polymer (RFP) surface over the touch surface using the fretting phenomenon. The frequency of vibrations is 500 Hz, which is more than the human haptic range. Also, use of RFP film may degrade display quality.

Nowadays, different materials are used in touch surfaces like Gorilla glass, Dragontrail glass [13-14] etc. A localized vibrotactile feedback generation strategy developed for one type of material with a certain type of actuator may not work for a different material. The development of a mechanical model considering the touch surface and the actuator configurations can aid in the convenient extension of methodology for one material and actuator configuration to another. In this research, we are proposing touch surfaces of bar/beam type with springs and dampers at the boundary. Mechanical models using the finite-element method for the beam-type surfaces are proposed with multiple electrostatic resonating actuators providing sinusoidal excitations in the human haptic perception range at different spatial positions. An analytical solution is presented to obtain vibrotactile response at different spatial locations of the touch surface for different actuator configurations excited by multiple excitation frequencies. The change in vibrotactile response for surfaces with different mechanical properties is demonstrated for similar actuator configurations and excitation frequencies. A variety of vibrotactile feedback generation with varying frequencies of excitation, actuator stiffness, actuator positions, and a number of actuators is presented. This allowed the exploration of optimal locations the number of actuators for desired vibrotactile acceleration magnitude and pattern for a prescribed spatial position on the surface. It is shown that a desired vibrotactile pattern across the touch surface, by changing excitation frequencies and actuator configuration, can be attained.

## MODELING AND ANALYSIS APPROACH

The touch surface considered in this study is a beam type or narrow touch bar surface with spring boundary conditions using electrostatic resonating actuators (ERAs). This touch bar interface configuration with ERAs represented as equivalent spring-damper boundary conditions is shown in figure 1. Electrostatic actuators are typically constructed in parallel plate configurations with one fixed electrode and one movable electrode connected to a spring. The movable electrode under the influence of an electric field can cause oscillations. Mason et al. [3] proposed a modified dual electrode ERA suitable for producing vibrotactile feedback in large touchscreen devices. In the dual-electrode ERA of [3], one grounded mass with a radial beam spring is inserted between the two electrodes to be energized. The electrodes are energized one at a time in an alternating manner to avoid interference of forces generated by lower and upper electrodes. The oscillations of the grounded mass can be utilized to excite touch surfaces through a central hole. The stiffness of the radial beam spring defining the actuator stiffness and equivalent damping is obtained based on the dynamic characterization of ERAs shown in [3]. It is assumed that the central hole of the actuator(s) are bolted to the touch bar and that the actuators' stiffness and damping get reflected at the boundary through the bolt(s). Here we take, $k_b$ as the actuator stiffness, $m_b$ as the mass of the bolt connecting each actuator to be touch bar, $c$ as the damping coefficient of the actuator with damping ratio $\zeta$, $a$ as the maximum amplitude of displacement input from the actuator. The ERAs provide sinusoidal displacements as base excitations which get transmitted to the touch bar through the. Figure 1 shows the configuration of the proposed touch bar with dual actuator excitations. The homogeneous mechanical properties of different materials considered to be used as a touch bar in this study are shown in table 1. The geometry of the touch bar is defined by length $L$, width $b$, and thickness $h$.

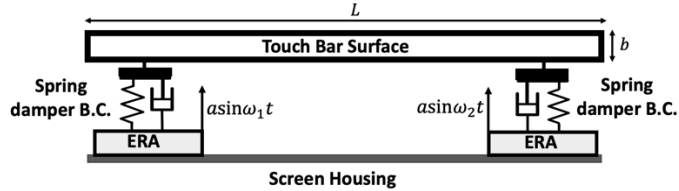

**FIGURE 1:** Proposed touch bar surface with ER actuators with spring-damper boundary conditions

**Table 1.** Mechanical properties of different materials of touch bar considered

| Material | Mechanical Property | |
|---|---|---|
| | E (GPa) | Density (kg/m$^3$) |
| Aluminum | 70 | 2700 |
| Dragontrail Glass | 74 | 2480 |
| Copper | 130 | 8960 |

**Finite element modeling of the touch bar system**

The touch bar surface is modeled as Euler-Bernoulli prismatic beam. Here, we consider $E$ be the uniform elastic modulus of the beam, $I$ be the uniform second moment of inertia, $A$ be the cross sectional area, $\rho$ be the density of the beam, $u(x,t)$ be the transverse displacement, $x$ be the spatial location, and $t$ be the time instant. The governing equation for a touch bar as an Euler-Bernoulli beam for forced vibration is given as [15],

$$EI\frac{\partial^4 u(x,t)}{\partial^4 x} + \rho A\frac{\partial^2 u(x,t)}{\partial^2 t} = f(x,t) \qquad (1)$$

We consider a weight function $w(x,t)$ ($\forall w = 0$ at $\Gamma_e$) with $\Gamma_e$ be the essential boundary, $\Gamma_t$ be the natural boundary, $\Gamma$ be the total boundary ($\Gamma \in \Gamma_e \cup \Gamma_t$), and $\Omega_x$ be the spatial domain. Therefore, the variational (weak) form of (1) is given as,

$$\int_{\Omega_x}\frac{d^2 w}{dx^2}EI\frac{\partial^2 u}{\partial x^2}dx + \int_{\Omega_x} w\rho A\frac{\partial^2 u}{\partial t^2}dx = \int_{\Omega_x} wfdx + w\bar{V}\Big|_\Gamma + \bar{M}\frac{dw}{dx}\Big|_\Gamma \qquad (2)$$

Here, $\bar{V}$ & $\bar{M}$ are shear force and moments applied at the boundary. Since $w(x)$ is a scalar, without loss of generality, we can rewrite equation (2) as,

$$\int_{\Omega_x} \frac{d^2 w^T}{dx^2} EI \frac{\partial^2 u}{\partial x^2} dx + \int_{\Omega_x} w^T \rho A \frac{\partial^2 u}{\partial t^2} dx = \int_{\Omega_x} w^T f dx + \left. w^T \overline{V} \right|_{\Gamma} + \left. \overline{M} \frac{dw^T}{dx} \right|_{\Gamma} \tag{3}$$

Now, we discretize the touch bar to 2 noded $n_{el}$ finite elements with $x_1^e$ & $x_2^e$ being the elemental boundaries. Therefore, equation (3) becomes,

$$\sum_{n=1}^{n_{el}} \left[ \int_{x_1^e}^{x_2^e} \frac{d^2 w^{e^T}}{dx^2} E^e I^e \frac{\partial^2 u^e}{\partial x^2} dx + \int_{x_1^e}^{x_2^e} w^{e^T} \rho^e A^e \frac{\partial^2 u^e}{\partial t^2} dx = \int_{x_1^e}^{x_2^e} w^{e^T} f^e dx + \left. w^{e^T} \overline{V}^e \right|_{\Gamma^e} + \left. \overline{M}^e \frac{dw^{e^T}}{dx} \right|_{\Gamma^e} \right] \tag{4}$$

We consider that the transverse displacement varies as a cubic polynomial across each element. The elemental transverse displacement can be given as,

$$u^e(x,t) = \alpha_0(t) + \alpha_1(t)x + \alpha_2(t)x^2 + \alpha_3(t)x^3 \tag{5}$$

If the elemental nodal displacements and slopes can be given as the vector $\mathbf{d}^e = [\ u_1^e(t)\ \ \theta_1^e(t)\ \ u_2^e(t)\ \ \theta_2^e(t)\ ]^T$ and $\mathbf{N}^e(x)$ is the elemental shape function, equation (5) can be rewritten as equation (6). Here, each element has two nodes and each node has two degrees of freedom (transverse displacement and slope).

$$u^e(x,t) = \mathbf{N}^e(x) \mathbf{d}^e(t) \tag{6}$$

Here, we are considering that there are only springs-dampers and a mass of bolts at the boundaries of the touch bar. Therefore, we have only the natural boundary conditions for each element, i.e., no moments are applied at the boundaries. So, the shear force acting on the elemental boundaries due to the proposed boundary conditions can be given as [16],

$$\left. \overline{V} \right|_{\Gamma_t^e} = n \mathbf{N}^e(x) \left[ m_b \ddot{\mathbf{d}}^e(t) + c \dot{\mathbf{d}}^e(t) + k_b \mathbf{d}^e(t) \right] \Big|_{\Gamma_{t_{LH}}^e}$$
$$- n \mathbf{N}^e(x) \left[ m_b \ddot{\mathbf{d}}^e(t) + c \dot{\mathbf{d}}^e(t) + k_b \mathbf{d}^e(t) \right] \Big|_{\Gamma_{t_{RH}}^e} \tag{7}$$

$$\left( \text{Here, n} = \begin{cases} -1 \text{ at the left end element boundary } \Gamma_{t_{LH}}^e \\ 1 \text{ at the right end element boundary } \Gamma_{t_{RH}}^e \end{cases} \right)$$

Considering the weighting function $w^e$ varies as a cubic polynomial within element $w^e(x) = \mathbf{N}^e(x) \mathbf{w}^e$. From equation (4) we obtain the elemental stiffness, mass, damping, external forces matrices as,

$$\mathbf{K}^e = \int_{x_1^e}^{x_2^e} \frac{d^2 \mathbf{N}^{e^T}}{dx^2} E^e I^e \frac{\partial^2 \mathbf{N}^{e^T}}{\partial x^2} dx - n \mathbf{N}^{e^T} \mathbf{N}^e k_b \Big|_{\Gamma_{t_{LH}}^e}$$
$$+ n \mathbf{N}^{e^T} \mathbf{N}^e k_b \Big|_{\Gamma_{t_{RH}}^e}$$
$$\Rightarrow \mathbf{K}^e = \int_{x_1^e}^{x_2^e} \mathbf{B}^{e^T} E^e I^e \mathbf{B}^e dx - n \mathbf{N}^{e^T} \mathbf{N}^e k_b \Big|_{\Gamma_{t_{LH}}^e}$$
$$+ n \mathbf{N}^{e^T} \mathbf{N}^e k_b \Big|_{\Gamma_{t_{RH}}^e} \tag{8}$$

$$\mathbf{M}^e = \int_{x_1^e}^{x_2^e} \mathbf{N}^{e^T} \rho^e A^e \mathbf{N}^e dx - n\mathbf{N}^{e^T} \mathbf{N}^e m_b \Big|_{\Gamma_{t_{LH}}^e} + n\mathbf{N}^{e^T} \mathbf{N}^e m_b \Big|_{\Gamma_{t_{RH}}^e} \qquad (9)$$

$$\mathbf{C}^e = n\mathbf{N}^{e^T} \mathbf{N}^e c \Big|_{\Gamma_{t_{LH}}^e} + n\mathbf{N}^{e^T} \mathbf{N}^e c \Big|_{\Gamma_{t_{RH}}^e} \qquad (10)$$

$$\mathbf{f}^e = \int_{x_1^e}^{x_2^e} \mathbf{N}^{e^T} f^e dx \qquad (11)$$

Now, scattering and adding the elemental stiffness, mass, damping, and external forces matrices of $n_{el}$ elements, we get the global stiffness matrix $\mathbf{K}$, mass matrix $\mathbf{M}$, and external forces matrix $\mathbf{f}$ defining the following differential equation of motion governing the dynamics of the touch bar system with spring-damper boundaries

$$\mathbf{M}\ddot{\mathbf{d}}(t) + \mathbf{C}\dot{\mathbf{d}}(t) + \mathbf{K}\mathbf{d}(t) = \mathbf{f}(t). \qquad (12)$$

**Analytical solution for vibrotactile feedback acceleration at spatial locations of the touch bar**

We have discretized the touch bar surface to $n=2(n_{el}+1)$ degree-of-freedom system governed by equation (12). If the system is excited using $m$ numbers of harmonic excitations, then the governing differential equation becomes,

$$\mathbf{M}\ddot{\mathbf{d}}(t) + \mathbf{C}\dot{\mathbf{d}}(t) + \mathbf{K}\mathbf{d}(t) = \sum_{i=1}^{m}\left(\mathbf{g}_i \sin \omega_i t + \mathbf{h}_i \cos \omega_i t\right) \qquad (13)$$

Here, $\mathbf{g}_i$, $\mathbf{h}_i$ are constant vectors with amplitudes of the harmonic excitations at $i$-th degree of freedom. Let the initial conditions be,

$$\mathbf{d}(0) = \mathbf{d}_0, \ \dot{\mathbf{d}}(0) = \mathbf{v}_0 \qquad (14)$$

The system response is the combination of free or homogeneous response $\mathbf{d}_H(t)$ and the forced or particular response $\mathbf{d}_P(t)$. $\mathbf{d}_H(t)$ is the solution of equation (15) and $\mathbf{d}_P(t)$ is the solution of equation (16). $\mathbf{d}_P(t)$ can be considered as the combination of all individual harmonic excitation responses $\mathbf{d}_{P_i}(t)$ as shown in equation (17).

$$\mathbf{M}\ddot{\mathbf{d}}_H(t) + \mathbf{C}\dot{\mathbf{d}}_H(t) + \mathbf{K}\mathbf{d}_H(t) = 0 \qquad (15)$$

$$\mathbf{M}\ddot{\mathbf{d}}_P(t) + \mathbf{C}\dot{\mathbf{d}}_P(t) + \mathbf{K}\mathbf{d}_P(t) = \sum_{i=1}^{m}\left(\mathbf{g}_i \sin \omega_i t + \mathbf{h}_i \cos \omega_i t\right) \qquad (16)$$

Therefore, the total response of the touch bar system with multiple harmonic excitations is given as,

$$\mathbf{d}(t) = \mathbf{d}_H(t) + \mathbf{d}_P(t) = \mathbf{d}_H(t) + \sum_{i=1}^{m}\mathbf{d}_{P_i}(t) \qquad (17)$$

We assume a solution of equation (16) of the form,

$$\mathbf{d}_P(t) = \sum_{i=1}^{m}\mathbf{d}_{P_i}(t) = \sum_{i=1}^{m}\left(\mathbf{p}_i \cos \omega_i t + \mathbf{q}_i \sin \omega_i t\right) \qquad (18)$$

$\mathbf{p}_i$, $\mathbf{q}_i$ are constant vectors representing the harmonic motion with the same frequency $\omega_i$ of the exciting force at $k$-th degree of freedon having different amplitude $p_k$ and/or $q_k$ of oscillations. Such a motion occurs at steady state when transient response of the system dies out. For $i^{th}$ excitation, the steady state response is of the form,

$$\mathbf{d}_{P_i}(t) = \mathbf{p}_i \cos \omega_i t + \mathbf{q}_i \sin \omega_i t \qquad (19)$$

Substituting equation (19) in equation (16) we obtain,

$$\begin{aligned}-\omega_i^2 \mathbf{M}\mathbf{p}_i \cos \omega_i t - \omega_i^2 \mathbf{M}\mathbf{q}_i \sin \omega_i t - \omega_i \mathbf{C}\mathbf{p}_i \sin \omega_i t \\ + \omega_i \mathbf{C}\mathbf{q}_i \cos \omega_i t + \mathbf{K}\mathbf{p}_i \cos \omega_i t + \mathbf{K}\mathbf{q}_i \sin \omega_i t = \\ \mathbf{g}_i \sin \omega_i t + \mathbf{h}_i \cos \omega_i t\end{aligned} \quad (20)$$

Comparing coefficients of both sides of equation(20), we obtain in matrix form,

$$\begin{bmatrix} \mathbf{K} - \omega_i^2 \mathbf{M} & \omega_i \mathbf{C} \\ -\omega_i \mathbf{C} & \mathbf{K} - \omega_i^2 \mathbf{M} \end{bmatrix} \begin{pmatrix} \mathbf{p}_i \\ \mathbf{q}_i \end{pmatrix} = \begin{pmatrix} \mathbf{h}_i \\ \mathbf{g}_i \end{pmatrix} \quad (21)$$

and hence, $\mathbf{p}_i$, $\mathbf{q}_i$ for a given excitation can be obtained whenever the left-hand side of equation (21) is not singular. The particular response due to $i^{th}$ harmonic excitation is thus obtained as in equation(19). The total particular response is obtained by solving for each one of the m harmonic excitations and superposition of all of them as in equation(18). The singularity of the left-hand side of equation (21) implies that the system is in resonance and there is no steady state response as the vibrations increase without bound.

For homogeneous solution of equation(15) is a quadratic eigenvalue problem. We can use numerical solvers for example MATLAB®'s *eig* to compute associated eigenvalues and eigen vectors after converting equation (15) to a suitable state space eigenvalue form

$$\left( \underbrace{\begin{pmatrix} \mathbf{0} & \mathbf{I} \\ -\mathbf{M}^{-1}\mathbf{K} & -\mathbf{M}^{-1}\mathbf{C} \end{pmatrix}}_{\mathbf{A}} - s \underbrace{\begin{pmatrix} \mathbf{I} & \mathbf{0} \\ \mathbf{0} & \mathbf{I} \end{pmatrix}}_{\mathbf{i}} \right) \underbrace{\begin{pmatrix} \mathbf{d}_H \\ s\mathbf{d}_H \end{pmatrix}}_{\mathbf{v}} = \begin{pmatrix} \mathbf{0} \\ \mathbf{0} \end{pmatrix}. \quad (22)$$

We solve the generalized eigenvalue problem (22) and obtain the eigen vector matrix $\mathbf{V}$. Let $(\mathbf{r}_1, \mathbf{r}_2, \cdots, \mathbf{r}_{2n})$ be the eigenvectors associated with complex eigenvalues $s_1, s_2, \cdots, s_{2n}$ respectively from the solution of (22). Considering that the 2n eigenvectors span the state space in a sense that $\mathbf{V}$ is invertible, we get the homogeneous response of the touch bar system as,

$$\mathbf{d}_H(t) = \sum_{j=1}^{2n} \varepsilon_j \mathbf{r}_j e^{s_j t} \text{ and } \mathbf{V} = \begin{bmatrix} \mathbf{r}_1 & \mathbf{r}_2 & \cdots & \mathbf{r}_{2n} \\ s_1\mathbf{r}_1 & s_2\mathbf{r}_2 & \cdots & s_{2n}\mathbf{r}_{2n} \end{bmatrix}. \quad (23)$$

By substituting (19) and (23) in (17), the complete solution for the touch bar system can be given as,

$$\mathbf{d}(t) = \sum_{i=1}^{m} (\mathbf{p}_i \cos \omega_i t + \mathbf{q} \sin \omega_i t) + \sum_{j=1}^{2n} \varepsilon_j \mathbf{r}_j e^{s_j t} \quad (24)$$

Using initial conditions from (14) in (24) we obtain the coefficients $\varepsilon_j$ (j=1,2,3,.....) and the complete solution of transverse displacements for the touch bar is obtained. For the zero initial condition the steady state amplitude of vibration can be obtained as,

$$\mathbf{d}(t) = \sum_{i=1}^{m} (\mathbf{p}_i \cos \omega_i t + \mathbf{q} \sin \omega_i t) \quad (25)$$

The vibrotactile response at different beam position can also be quantified using steady state acceleration value at different spatial locations of the beam. The steady state accelerations at different nodal positions of the touch bar hence can be given by,

$$\mathbf{a}(t) = \ddot{\mathbf{d}}_P(t) = -\sum_{i=1}^{m} \omega_i^2 (\mathbf{p}_i \cos \omega_i t + \mathbf{q} \sin \omega_i t) \quad (26)$$

Therefore, the absolute peak steady-state amplitude of acceleration at any degree of freedom $k$ of the touch bar can be obtained as given as,

$$\left(\alpha_p\right)_k = \sum_{i=1}^{m} \omega_i^2 \sqrt{\left(p_k\right)_i^2 + \left(q_k\right)_i^2} \quad (27)$$

**RESULTS AND DISCUSSION**

The finite element (FE) model of the touch bar surface is formulated in MATLAB® as delineated in the previous section, which can accommodate any change in the touch surface's mechanical properties, actuators' stiffness, actuators' positions, and actuators' actuators' damping. To validate the FE model, an aluminum touch bar is considered simply supported at both ends (pinned-pinned). The analytical natural frequencies of a pinned-pinned prismatic beam are given by [16],

$$\omega_n = (n\pi)^2 \sqrt{\frac{EI}{\rho A L^2}}, \text{ for } n = 1, 2, 3, \ldots \quad (28)$$

The natural frequencies of the aluminum touch bar in pinned-pinned configuration are also computed from the FE model using 62 degrees of freedom. The computed natural frequencies with $E = 70$ GPa, $L = 12$ inch, $b=0.984$ inch, and $h=0.03937$ inch for the pinned-pinned case are summarized in table 2. The results in table 2 validates the accuracy of the finite element model of the touch bar system.

**Table 2.** Natural Frequencies of pinned-pinned aluminum touch bar using 62 DoF FE model

| Mode Shape | Natural Frequencies (Hz) | | % error |
|---|---|---|---|
| | Analytical | FE Model | |
| 1 | 24.8197 | 24.8197 | 0 |
| 2 | 99.2788 | 99.2790 | 0.000201 |
| 3 | 223.3774 | 223.3789 | 0.000672 |
| 4 | 397.1154 | 397.1238 | 0.002115 |
| 5 | 620.4928 | 620.5249 | 0.005173 |

We have proposed analytical method in the previous section to obtain vibration response of the touch bar system for harmonic excitation. Apart from the benefits of close form solution obtained from the analytical method, it also aids in significant reduction in computation time compared to numerical solvers like ODE45 available in MATLAB®. To validate our analytical solution method, FE model of the aluminum touch bar in pinned-pinned configuration is considered. Dual sinusoidal excitation of maximum amplitude 2 N at locations 0.1L and 0.9L across the length of the beam are considered. The FE model is simulated with 22 degrees of freedom and the transverse displacements at two randomly picked nodes are compared. The comparison is shown in figure 3.

The results in figure 2 validate the proposed analytical solution method for the computation of vibration response of the touch bar system by comparing it with the numerical solution obtained using MATLAB's ODE45 solver.

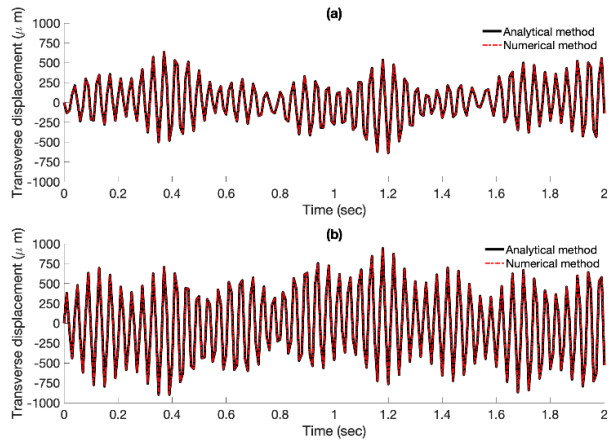

**FIGURE 2:** Comparison of transverse displacements calculated for the fe model of the touch bar with pinned-pinned configuration at a) Node #2, b) Node #7 using the proposed analytical method and numerical method in MATLAB®.

**Effect of the Material Properties on Touch Bar**

Touch bar surfaces made up of three different materials are considered: aluminum, Dragontrail glass, and copper. The vibrotactile responses of the touch surfaces consisting of the three different materials are considered for single actuator excitation and dual actuator excitation. The actuator stiffness is taken to be 16.18 kN/m for subsequent simulation unless specified otherwise. A small damping ratio of 0.02 and a maximum amplitude of displacement output of 0.04125 mm are considered based on the construction and operation parameters in [7]. The actuators provide sinusoidal displacements of specified frequencies. For single actuator excitation, the actuator is placed at 0.03L length of the touch bar. For dual actuator excitation, the first actuator is placed at 0.03L and the second actuator is placed at 0.97L across the length of the touch bar.

From figure 3, we observe that the copper touch bar produces maximum peak acceleration close to three times that of that produced by Dragontrail glass and aluminum touch bar. Near 0.25L length of the touch bar, the peak acceleration graph of copper just crosses a crest, and that of aluminum and Dragontrail glass approaches their respective troughs. From figure 4, it is observed that the Dragontrail glass exhibits the highest peak acceleration. The peak acceleration curve for the copper touch bar approaches troughs at multiple places near the crests of the peak acceleration graphs for the aluminum and Dragontrail glass and vice versa. Therefore, we have observed that mechanical properties of the touch surface impact the vibrotactile response of touch surfaces significantly. Localizable vibrotactile feedback strategies developed for touch surfaces of a particular material may need modifications for a similar surface with a different material. A mechanical model as developed in this research can be instrumental in making such necessary modifications. Otherwise, physical testing and calibration may be needed which involves significant man hours and cost.

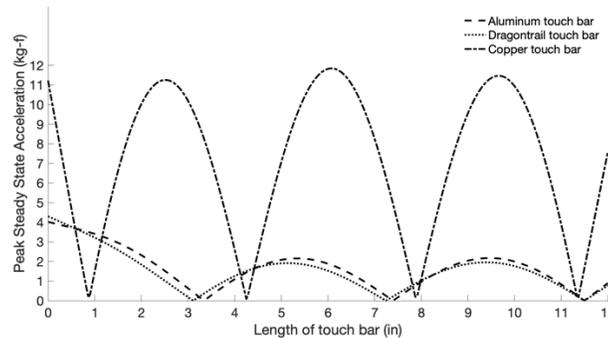

**FIGURE 3:** Peak accelerations across the length of the touch bar for single actuator excitation at 205 hz

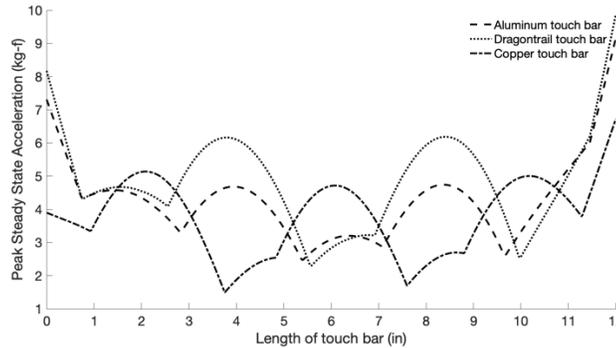

**FIGURE 4:** Peak accelerations across the length of the touch bar for dual actuator excitation (Left hand actuator at 179 hz and right-hand actuator at 157 hz)

**Effect of number of actuators and their positions**

To study the effect of the number of actuators and their positions, we need to look at the peak acceleration amplitudes offered with sinusoidal excitations of frequencies at which humans are most sensitive to vibrotactile stimuli. Physically testing the touch bar for different cases is not practical for this task. With the use of the proposed FE model, we can statistically analyze the vibrotactile response of the touch surface for different actuators and their positions. To demonstrate, we have considered aluminum touch bar surface. For brevity, we consider three cases: single actuator, dual actuators, and triple actuators. The configurations of the positions for the dual actuator case are given in table 3. The configurations of the positions for the triple actuator case are given in table 4. We consider excitations with a range of frequencies with full displacement amplitude of the ERAs in the human haptic range (150-250 Hz). For dual and triple actuator configurations, excitations of both similar and different frequencies at all the two or three actuator positions are considered.

Figure 5 shows the boxplots and violin plots together for different actuator positions across the touch bar for the case of single excitation. For each location of the actuator, the touch bar is excited with the same range of frequencies between 150-250 Hz, and the cumulative observations for peak accelerations obtained across the length of the touch bar are plotted. We observe that the median peak acceleration for single actuator excitation is less than 1 g for all the positions except the 0.16L position. The density of peak accelerations is also seen to be below 1 g peak acceleration mostly, except at the 0.16L position. Also, the actuator at position 0.16L is capable of a wider band of peak accelerations from 1.5 g to ~ 4 g. Therefore, for the considered positions of the single actuator, the optimum position is found to be at 0.16L across the length of the touch bar.

**Table 3.** Actuator position configurations for dual actuator case

| Configuration | Position across the length of the touch bar | |
| --- | --- | --- |
| | Left hand side actuator | Right hand side actuator |
| 1 | 0 | L |
| 2 | 0.03L | 0.97L |
| 3 | 0.16L | 0.84L |
| 4 | 0.33L | 0.67L |
| 5 | 0.49L | 0.51L |

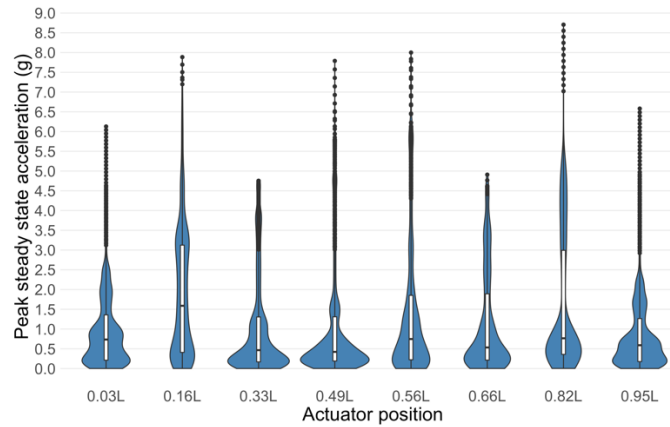

**FIGURE 5:** Cumulative peak acceleration magnitudes at different positions of the single actuator for a range of excitation frequencies (150-250 hz)

Figure 6 shows the boxplots and violin plots together for different actuator positions across the touch bar for dual actuator excitation. For configurations of positions of the two actuators (as in table 3), the touch bar is excited with the same range of frequencies between 150-250 Hz and the cumulative observations for peak accelerations obtained across the length of the touch bar are plotted. The median peak acceleration is found to be less than 2 g for all the configurations except configuration 3 where the median peak acceleration is slightly above 3 g. The densities of the peak accelerations for all the configurations are found to be concentrated below 3 g except for configuration 3. Also, configuration 3 can generate peak accelerations from 3 g to 7 g at a much greater density compared to the rest of the configurations. Therefore, for the position configurations considered for dual actuator excitations, the optimum configuration is found to be configuration 3.

**Table 4.** Actuator position configurations for triple actuator case

| Configuration | Position across the length of the touch bar | | |
| --- | --- | --- | --- |
| | Left hand side actuator | Middle actuator | Right hand side actuator |
| 1 | 0 | 0.49L | L |
| 2 | 0.03L | 0.49L | 0.97L |
| 3 | 0.16L | 0.49L | 0.84L |
| 4 | 0.33L | 0.49L | 0.67L |
| 5 | 0.43L | 0.49L | 0.57L |

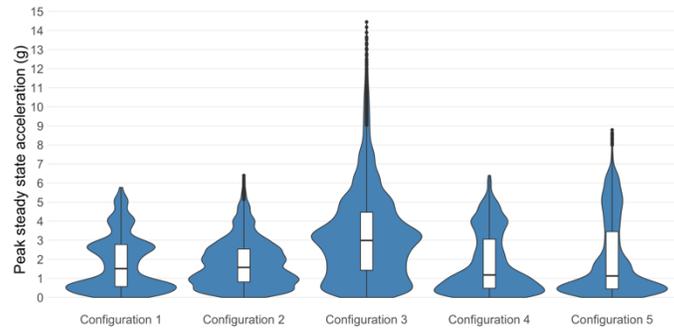

**FIGURE 6:** Cumulative peak acceleration magnitudes at different position configurations (As in table 3) of dual actuators for a range of excitation frequencies (150-250 hz)

Figure 7 shows the boxplots and violin plots together for different actuator positions across the touch bar for triple actuator excitation. For configurations of positions of the three actuators (as in table 4), the touch bar is excited with the same range of frequencies between 150-250 Hz, and the cumulative observations for peak accelerations obtained across the length of the touch bar are plotted. The median peak acceleration is found to be more than 3 g for all the configurations except configuration 5. The highest median peak acceleration is found for configuration 1 as close to 6 kg-f. Also, configuration 1 provides a wider range of peak accelerations than the rest of the configurations. Therefore, for the position configurations considered for triple actuator excitations, the optimum configuration is found to be configuration 1. Configuration 1 can provide peak accelerations in the range 3-15 g. Configurations 2 and 3 can also provide reasonable peak accelerations from 3-12 g as observed from the densities of figure 8.

From the above analysis, we have observed that peak acceleration across the touch bar depends on the placement of the actuators. In general, we can infer that an increasing number of actuators tend to increase the magnitude of peak accelerations across the length of the touch bar.

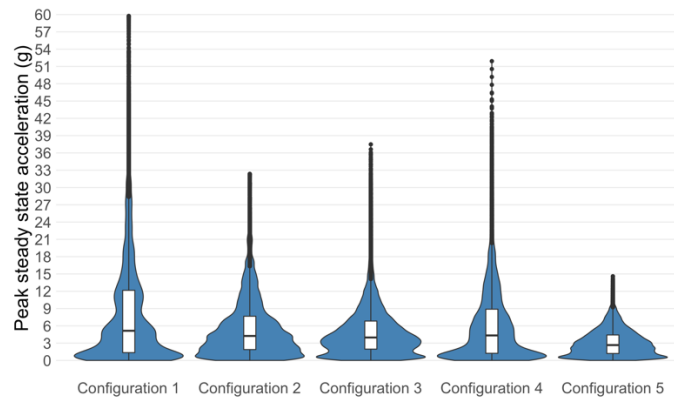

**FIGURE 7:** Cumulative peak acceleration magnitudes at different position configurations (As in table 4) of three actuators for a range of excitation frequencies (150-250 hz)

In order to understand variation in localized vibrotactile feedback at a desired location of the touch bar, acceleration at a fixed location (0.59L) on the bar is considered. The influence of the positions of the two actuators across the length of the touch bar on the variability of peak acceleration at the position $0.59L$ is observed. Different configurations of the two actuators as given in table 3 are considered for this analysis.

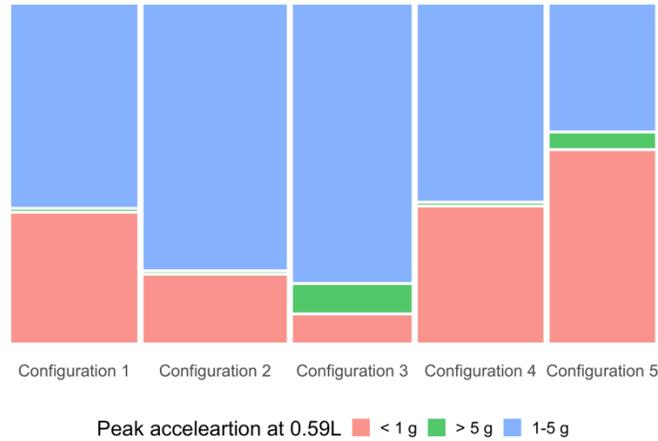

**FIGURE 8:** Mosaic plot depicting the fractions of peak accelerations < 1 g, > 5 g, and 1-5 g at different position configurations of dual actuators (As in table 3)

The touch bar is excited with a range of frequencies in the range 150-250 Hz and the peak accelerations at location 0.59L are captured for each position configuration. Figure 8 shows the mosaic plot depicting the fractions of peak accelerations at 0.59L position less than 1 g, greater than 5 g and between 1-5 g for each position configuration. For the dual actuator position configuration 3, the fraction of peak acceleration greater than 5 g (19% of the total responses obtained) and the fraction of peak acceleration between 1-5 g (61% of the total responses obtained) is the highest among the configurations considered. Also, this configuration provides the least fraction of peak acceleration less than 1 g (20% of the total responses obtained). Therefore, position configuration 3 for dual actuator excitation can be considered optimal (out of the considered configurations) for obtaining variable vibrotactile feedback at the location 0.59L of the touch bar. In this way, we can seek to obtain a desirable actuator configuration for a localizable vibrotactile feedback at a desired zone of the touch surface for different number of actuators.

**Effect of the frequency and stiffness of actuators**

We try to observe the vibrotactile response of the FE model of the touch using different frequencies for aluminum. For a single actuator positioned at 0.16L of the touch bar, we excited the touch bar with different frequencies in the human perceptible haptic range. For some frequencies of excitations, multiple nodes (with insignificant peak acceleration < 1g) were observed. By switching between frequencies, many of the nodes can be nullified. However, for single excitations, most cases are found such that even with switching between frequencies, some nodes can't be nullified. Figure 9 shows such a scenario for single excitation. Here, the node at 6 inch length of the bar due to 150 Hz excitation are nullified by switching to 250 Hz excitation. However, the nodes between 10-11 inch length of the bar could not be nullified.

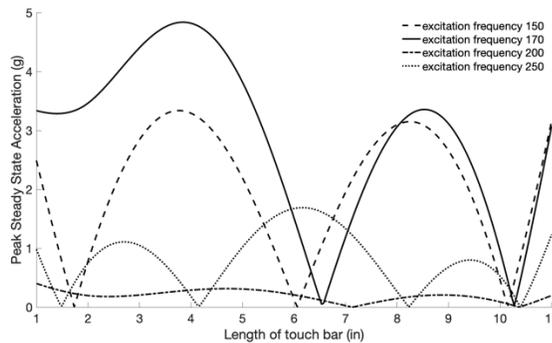

**FIGURE 9:** Variation of peak accelerations across the length of the touch bar for single excitation at 0.16L of the touch bar with varying frequencies

We excited the touch bar with different frequencies in the human perceptible haptic range for single dual actuators positioned at 0.16L and 0.84L of the touch bar. In the dual actuator configuration, the nodes appearing for a particular frequency can be nullified using other sets of frequencies. Therefore, we can achieve localizable vibrotactile feedback by switching between sets of frequencies. Figure 10 shows such a scenario for dual actuator excitation. The nodes between 6-7 inch length of the bar due to left actuator frequency 170 Hz and right actuator frequency 200 Hz are nullified by switching to left actuator frequency 170 Hz and right actuator frequency 230 Hz. At every point on the surface, the net peak acceleration is more than 3 g.

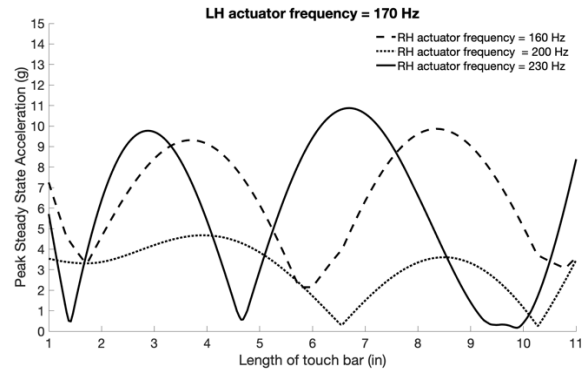

**FIGURE 10:** Variation of peak accelerations across the length of the touch bar for dual excitation with varying frequencies

For triple actuator excitation with the configuration positioned as per configuration 2 of table 4, we observe similar results like dual actuator excitation. Figure 11 shows such a scenario for triple actuator excitation. The nodes appearing between 10-11 inch length of the bar for one set of frequencies can be nullified by switching to other frequency sets. We also observe a near node situation between 3-4 inch bar length at a particular set of frequencies. The other sets of frequencies also nullify those near node points. At every point on the surface, the net peak acceleration is more than 5 g.

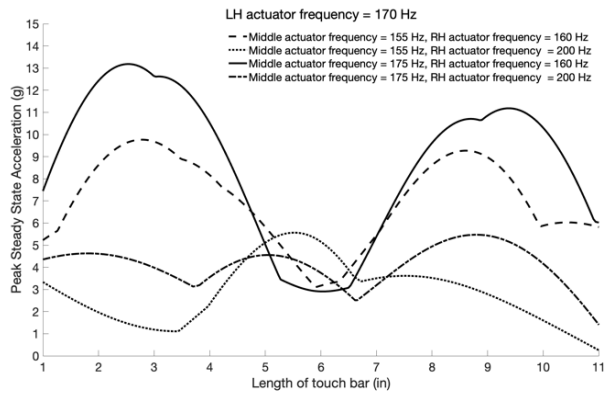

**FIGURE 11:** Variation of peak accelerations across the length of the touch bar for triple excitation with varying frequencies

We observed the variation of vibrotactile response across the touch bar for changing actuator stiffness. Figure 12 shows an observation of change in peak accelerations with different actuator stiffness values for dual excitation as per configuration 3 of table 3. We observe that the magnitude of peak acceleration across the length of the bar increases with the increase in actuator stiffness for the majority of the actuator locations. We also observe a slight variation in node points; however not that significant. We can adjust the actuator stiffness at different actuator numbers and positions to bump up the peak accelerations across the touch bar. Figure 13 shows the single actuator case in figure 9, but with an actuator stiffness of 25 kN/m. In figure 13, the nodes between 10-11 inches of the touch bar get nullified due to increased stiffness.

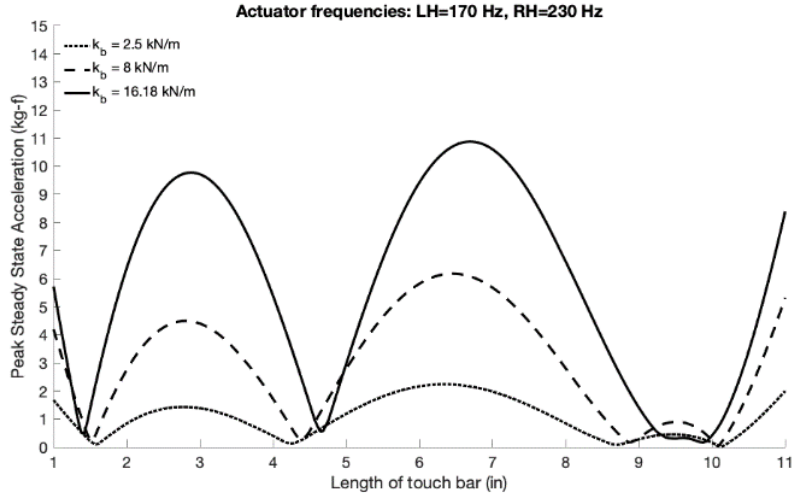

**FIGURE 12:** Variation of peak accelerations across the length of the touch bar for dual excitation with varying actuator stiffness

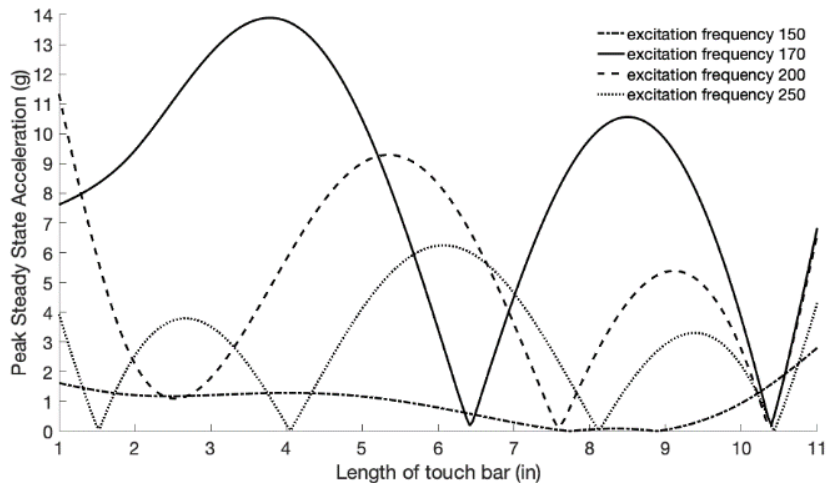

**FIGURE 13:** Variation of peak accelerations across the length of the touch bar for single excitation at 0.16L of the touch bar with varying frequencies with actuator stiffness of 25 kN/m

These results suggest that a wide variety of localizable vibrotactile feedback can be achieved using by varying the frequencies of excitation and actuator stiffness.

**CONCLUSION**

Mechanical model of a narrow touch bar type display has been developed using finite element method incorporating spring-damper boundary conditions of attached electrostatic actuators. An analytical solution method is proposed for finding the vibrotactile response of the touch bar for different actuator configurations with multiple harmonic base excitations. It has been shown that the material properties of the touch surface result in different vibrotactile feedback for similar actuator configurations and excitations. The simulation presented here using the FE model of touch bar display, can aid in selection of suitable number of actuators and their positions for vibrotactile feedback of desired magnitude across the length of the touch bar and at specified locations. Changes in actuator numbers and their position result in a variety of vibrotactile feedback across the touch bar. This study also demonstrates that a wide variety of vibrotactile feedback can be generated across the touch bar using at least two electrostatic actuators by varying the excitation frequencies. The variability of vibrotactile feedback due to changes in actuator stiffness is also demonstrated. It is shown that change in actuator stiffness coupled with change in excitation frequencies can generate a wide variety of localizable vibrotactile feedback across the touch bar. Subsequent future work will include experimental verification of bar type displays with multiple actuators and extension of this study to a 2-dimensional plate type touch surfaces.